\documentclass[aip,apl,twocolumn,floatfix,groupedaddress]{revtex4}
\usepackage{amsmath}
\usepackage{amssymb}
\usepackage[dvips,dvipdfm]{graphicx}
\usepackage{bm}
\begin{document}

\title{Electron tunneling through graphene-based
double barriers driven by a periodic potential}
\author{H. P. Ojeda-Collado}
\author{C. Rodr\'{\i}guez-Castellanos}
\affiliation{Department of Theoretical Physics, Faculty of Physics University of Havana,
San L\'{a}zaro y L, Vedado, La Habana, Cuba, CP 10400}
\author{}
\affiliation{}
\date{\today }

\begin{abstract}
Photon-assisted charge transport through a double barrier structure
under a time periodic field in graphene is studied. Within the
framework of the Floquet formalism and using the transfer matrix
method, the transmission probabilities for the central band and
sidebands are calculated. A critical phase difference between the
harmonic potentials at the barriers, which cancels transmission
through the inelastic sidebands due to quantum interference is
found. This phenomenon could be of help to
design graphene based filters and high-frequency radiation
detectors. Quenching of resonant tunneling by the harmonic field applied to the barriers or the well is also discussed.
\end{abstract}

\maketitle

Quantum transport in periodically driven mesoscopic systems is an
important subject not only of academic value but also for device
applications. The interest on time-dependent excitations by
electromagnetic fields has been increasing and many interesting
phenomena have been investigated~\cite{1,2}. Early studies of Dayem and
Martin~\cite{3} provided the evidence of photon assisted tunneling (PAT) in experiments on
superconducting films under microwave fields, and subsequently Tien
and Gordon theoretically justified this observation~\cite{4}. After
this, electron transport through various types of time-oscillating
potential regions has been studied in
semiconductor nanostructures~\cite{5,6,7,8} aiming to the fabrication of new devices.\\
A few years ago, graphene, a two dimensional honeycomb crystal of
carbon atoms, was fabricated by Novoselov \begin{itshape}et
al.\end{itshape}~\cite{9}, leading to increasing attention to both
the fundamental physics and potential applications of the new
material. In graphene, low-energy quasiparticle excitations near the
edges of the hexagonal Brillouin zone (Dirac points) obey a gapless
linear dispersion law, and their motion can be described by a
two-dimensional Dirac equation for masless particles. The presence
of such Dirac-like quasiparticles leads to Klein tunneling and other unusual electronic
properties~\cite{9,10,11,12,13,14}. For graphene-based PAT devices it is
essential to consider transport of charge carriers in graphene
through time-harmonic potentials. PAT through single barrier in
monolayer graphene have been discussed in Ref.~\cite{15}. This
letter is an extension for double barrier structures (DBS) taking
into account quantum interference between photon assisted processes.
At normal incidence PAT is analyzed aiming to get a direct insight
into the quantum interference and analytical condition for the
suppression of the transmission through the inelastic sidebands is
obtained. We also find that quantum interference makes a
significant contribution to the total transmission probability at
non-normal incidence and could be useful in
graphene-based PAT devices.\\
A single electron transmitting through a monolayer graphene-based
double barrier driven by a harmonic potential is considered. This
structure can be fabricated by applying a local top gate voltage
and a small ac signal to graphene. At low energy, and close to the
Dirac points (K and K') electrons are described by a massless Dirac
Hamiltonian:
\begin{equation}
\hat{H}=v_{f}\hat{\vec{\sigma}}\cdot\hat{\vec{p}}+\hat{V}(x,t)\label{hamil}\end{equation}
with\[ \hat{V}(x,t)=\hat{I}\left[V_{0}+V(x)cos\left(\omega
t+\delta\,\vartheta\left(x\right)\,\right)\right] \]
\begin{equation}
\times\vartheta\left(|x|-\frac{d}{2}\right)\vartheta\left(\frac{L}{2}-|x|\right)\label{potential}
\end{equation}
where the Fermi velocity $v_{f}\thickapprox10^{6}m/s$,
$\hat{\vec{\sigma}}=(\hat{\sigma_{x}},\hat{\sigma_{y}})$ is the 2D
vector of Pauli matrices and
$\hat{\vec{p}}\rightarrow-i\hbar\vec{\nabla}$ is the 2D momentum
operator. The barriers of height
$V_{0}$ are oscillating with frequency $\omega$, phase difference
$\delta$, and different amplitude: $V(x)=V_{1}$ if $x<0$ or
$V(x)=V_{3}$ for $x>0$. This simple device is shown schematically in Fig. 1.
\begin{figure}[tbp]
\hspace*{-.2cm}\includegraphics[scale=0.56]{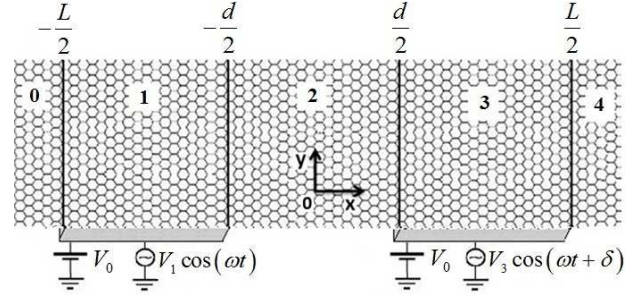}
\caption{Schematic oscillating potential of the DBS in graphene.}
\label{fig1}
\end{figure}
The barriers and well widths are $L_{1}=(L-d)/2$ and $L_{2}=d$ respectively,
$\vartheta(x)$ is the Heaviside function and $\hat{I}$ is unit
matrix. The DBS is infinite and homogeneous along the y-direction,
resulting in the conservation of the y-component of momentum.
Incident electrons propagate from left and pass five regions
denoted by $0,$ $1,$ $2,$ $3,$ $4.$ The solution of Dirac-like
equation with hamiltonian (\ref{hamil}) for given quasienergy $E$
and y-component of momentum $k_{y}$ can be written as a Floquet
state:\[
\psi^{r}(x,y)=\frac{e^{ik_{y}y}}{\sqrt{2}}\sum_{m,n=-\infty}^{+\infty}\left[a_{n}^{r}\left(\begin{array}{c}
1\\
s_{n}^{r}e^{i\varphi_{n}^{r}}\end{array}\right)e^{ik_{n}^{r}\left(x-x_{r}\right)}\right.\]
\begin{equation}
 \left.+b_{n}^{r}\left(\begin{array}{c}
1\\
-s_{n}^{r}e^{-i\varphi_{n}^{r}}\end{array}\right)e^{-ik_{n}^{r}\left(x-x_{r}\right)}\right]
\end{equation}
\[\times J_{m-n}\left(\frac{V_{r}}{\hbar\omega}\right)e^{-i(m-n)\delta_{r}}e^{-i(E+m\hbar\omega)t/\hbar}\]
where the script $r=0,1,2,3,4$ indicates the region, $x_{r}$ are the coordinates of the boundaries: $x_{0}\equiv
x_{1}=-L/2,$ $x_{2}=-d/2,$ $x_{3}=d/2,$ $x_{4}=L/2,$ (see Fig. 1)
and $J_{n}$ is the Bessel function of the first kind. In barrier
regions $(r=1,3)$:
\begin{equation}
\begin{array}{c}
 s_{n}^{r}=s'_{n}=sgn(E+n\hbar\omega-V_{0}),\\
k_{n}^{r}=q_{n}=\sqrt{\left(\frac{E-V_{0}+n\hbar\omega}{\hbar
v_{f}}\right)^{2}-k_{y}^{2}},\\
\varphi_{n}^{r}=\theta_{n}=\arctan(k_{y}/q_{n}).
\end{array}
\end{equation}
On the other hand for $r=0,2,4$:
\begin{equation}
\begin{array}{c}
 s_{n}^{r}=s_{n}=sgn(E+n\hbar\omega),\\
k_{n}^{r}=k_{n}=\sqrt{\left(\frac{E+n\hbar\omega}{\hbar
v_{f}}\right)^{2}-k_{y}^{2}},\\
\varphi_{n}^{r}=\phi_{n}=\arctan(k_{y}/k_{n})
\end{array}
\end{equation}
and $J_{m-n}(V_{r}/\hbar\omega)=\delta_{m,n}$ because in this case,
the modulation amplitude is $V_{r}=0$. The phase difference $\delta_{r}=\delta$
only for $r=3$, elsewhere is zero. As Dirac electrons pass through a
region subjected to time-harmonic potentials, transitions from the
central band to sidebands (channels) at energies $E\pm m\hbar\omega$
$\left(m=0,1,2,...\right)$ occur as electrons exchange
energy quanta with the oscillating
field.\\
The wave function is continuous at the boundaries, and the
continuity condition can be expressed as:
\begin{equation} \left(\begin{array}{c}
\mathbb{A}^{0}\\
\mathbb{B}^{0}\end{array}\right)=\left(\begin{array}{cc}
\mathbb{K}_{11} & \mathbb{K}_{12}\\
\mathbb{K}_{21} & \mathbb{K}_{22}\end{array}\right)\left(\begin{array}{c}
\mathbb{A}^{4}\\
\mathbb{B}^{4}\end{array}\right)=\mathbb{K}\left(\begin{array}{c}
\mathbb{A}^{4}\\
\mathbb{B}^{4}\end{array}\right),\end{equation}
where the total transfer matrix $\mathbb{K}=\mathbb{K}(0,1)\mathbb{\cdot K}(1,2)\cdot\mathbb{K}(2,3)\cdot\mathbb{K}(3,4)$,
and $\mathbb{K}(r,r+1)$ $\left(r=0,1,2,3\right)$ are transfer matrices
that couple the wave function in the $r$-th region to the wave function
in the $\left(r+1\right)$-th region. We assume an electron propagating from left to right, then, $\mathbb{A}^{0}=\left\{ \delta_{n,0}\right\} $ and
$\mathbb{B}^{4}$ is the null vector, whereas $\mathbb{A}^{4}=\left\{ a_{n}^{4}\right\}$ and
$\mathbb{B}^{0}=\left\{ b_{n}^{0}\right\}$ are the coefficient vectors of transmitting waves and of reflecting waves respectively. The
total transmission probability for quasienergy $E$ is
$T=\sum_{n=-\infty}^{+\infty}T_{n},$ where
\begin{equation}
T_{n}=\frac{s_{n}cos(\phi_{n})}{s_{0}cos(\phi_{0})}|a_{n}^{4}|^{2}\label{channeltra}
\end{equation}
is the probability of the scattering for an
electron with incident quasienergy $E$ in region $0$ into the
sideband with quasienergy $E+n\hbar\omega$ in region $4$. The
minimum number $N$ of sidebands that need to be considered is
determined by the strength of the oscillation,
$N>max(V_{1}/\hbar\omega,V_{3}/\hbar\omega)$~\cite{8,15}, and the
infinite series for $T$ can be truncated to consider a finite number
of terms starting from $-N$ up to $N$.\\
Numerical calculations have been made for barriers and well of the same width: $d=50$ $nm$, $L=150$ $nm$, $V_{0}=200$ $meV$, $\omega=5\times10^{12}$ $Hz$
and $\lambda=50$ $nm$ for the wavelength of incident electrons.\\
\textbf{Normal incidence.}\\
The dependence of the transmission probability for the central band $\left(T_{0}\right)$
on $\alpha_{1}=V_{1}/\hbar\omega$ and $\alpha_{3}=V_{3}/\hbar\omega$
for normally incident electrons is shown in Fig. 2 for two values
of the phase difference $\delta.$
\begin{figure}[tbp]
\hspace*{-.7cm}\includegraphics[scale=0.45]{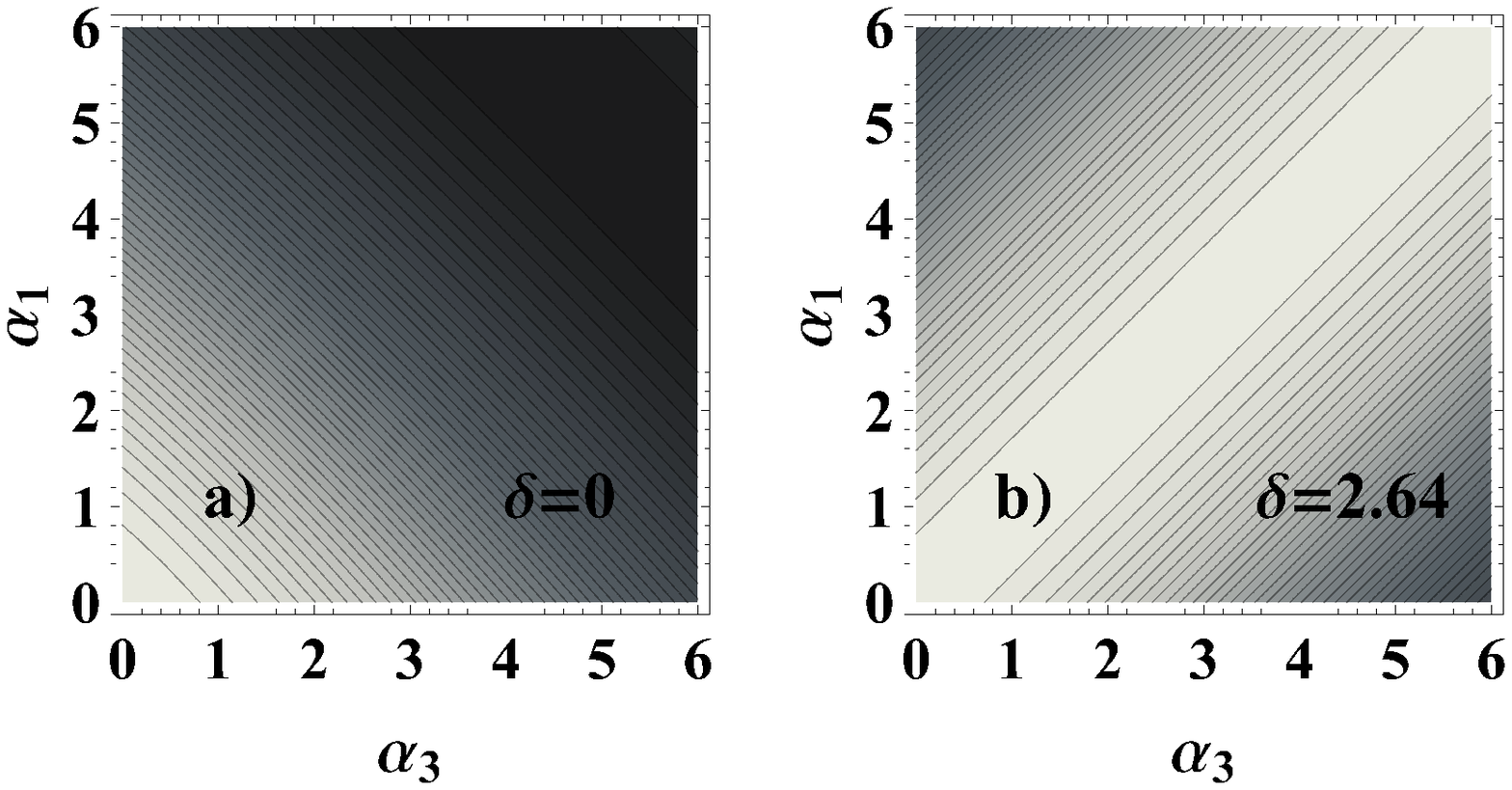}
\includegraphics[scale=0.55]{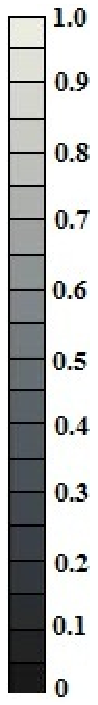}
\caption{Transmission probability for central band as a function
of $\alpha_{1}$ and $\alpha_{3}$ at $\phi_{0}=0$ for two value of $\delta.$}
\end{figure}
For small values of $\alpha_{1}$ and $\alpha_{3}$ the transmission
is perfect and take place only through the elastic band, because the
oscillating barrier can be treated approximated by a static one, and
for normal incidence, perfect transmission (Klein tunneling) through
static single~\cite{13} and double barriers has been
obtained~\cite{16}. With increasing $\alpha_{1}$ and $\alpha_{3}$,
higher and lower sidebands become important since electrons can
exchange a large number of photons with the time-periodic field,
decreasing the transmission probability through the central band
$\left(T_{0}\right)$ because these probabilities are now spread.
However, numerical calculations show the existence of a critical
value $\delta_{c}=2.64$ of the phase difference, at which the
transmission probability is exclusively through the central band
even for large $\alpha_{1}$ and $\alpha_{3}$ whenever
$\left|\alpha_{1}-\alpha_{3}\right|<1$. Thus, the quasienergy of the
transmitted electrons is sensitive to the phase difference, because
tuning $\alpha_{1},\alpha_{3}$ and $\delta$
could eliminate inelastic sidebands doing the function of an energy filter.\\
In order to understand the phenomenon behind this result, an
analytical expression for $T_{0}$ at normal incidence is derived.
Let us take the barriers with the same modulation amplitudes $\alpha_{1}=\alpha_{3}$. To simplify our calculation and get some direct insight
into the interference effect~\cite{17}, we take $\alpha_{1}$ as a
perturbation, and include the lowest order corrections up to
$\left(V_{1}/\hbar\omega\right)^{2},$ retaining only the terms
corresponding to the central band and first sidebands. Hence, the
probability for tunneling through the central band can be written as:
\begin{equation}
T_{0}=\left|a_{0}^{4}\right|^{2}=\frac{\gamma+\frac{1}{8}\left(\frac{\hbar\omega}{V_{1}}\right)^{2}}{1+\frac{1}{8}\left(\frac{\hbar\omega}{V_{1}}\right)^{2}}.\label{t0}
\end{equation}
The term $\gamma=\left(zp^{\star}+z^{\star}p\right)/8$ contains all the interference effects. The complex values $z$ and $p$ are given
by:
\begin{equation}
 \begin{array}{c}
  z=2e^{i\left(\delta+k_{0}L_{2}+L_{1}\left(q_{-1}+q_{0}+q_{1}\right)\right)}\left[e^{iq_{-1}L_{1}}+e^{iq_{1}L_{1}}\right]\\
+e^{i\left(k_{-1}L_{2}+2L_{1}\left(q_{-1}+q_{0}+q_{1}\right)\right)}\left[e^{-2iq_{-1}L_{1}}+e^{-2iq_{0}L_{1}}\right]\\
-2e^{i\left(k_{-1}L_{2}+L_{1}\left(q_{-1}+q_{0}+2q_{1}\right)\right)}\\
+e^{i\left(2\delta+k_{1}L_{2}+2L_{1}\left(q_{-1}+q_{0}+q_{1}\right)\right)}\left[e^{-2iq_{1}L_{1}}+e^{-2iq_{0}L_{1}}\right]\\
-2e^{i\left(2\delta+k_{1}L_{2}+L_{1}\left(2q_{-1}+q_{0}+q_{1}\right)\right)}
 \end{array}
\end{equation}
and
\begin{equation}
p=e^{i\left(\delta+k_{0}L_{2}+2L_{1}\left(q_{-1}+q_{1}\right)\right)}.
\end{equation}

Thus, when $\gamma=1$ the suppression of inelastic sidebands is obtained
under the condition: \begin{equation}
\delta_{c}\pm\left(k_{\pm1}-k_{0}\right)L_{2}\pm\left(q_{0}-q_{\pm1}\right)L_{1}=\pi.\label{condinterferencia}\end{equation}
The left-hand side of (\ref{condinterferencia}) is the phase
difference between two amplitudes. The first one corresponds to
particles that propagate through the DBS and absorb (emit) the
energy quantum $\hbar\omega$ at the first barrier, whereas the
second one corresponds to the same propagation with absorption
(emission) at the second barrier. Noteworthy that
$\pm\left(k_{\pm1}-k_{0}\right)L_{2}$, is the phase difference
between the amplitude corresponding to particles that absorb ($+$) or
emit ($-$) a photon close to the left barrier and traverse the well
with energy $E\pm \hbar\omega$ and the amplitude corresponding to
particles that absorb (emit) a photon close to the right barrier and
traverse the well with energy $E.$ In the same way,
$\pm\left(q_{0}-q_{\pm1}\right)L_{1}$ is the spatial phase
difference between the above mentioned amplitudes through the barrier
region. Therefore, the suppression of transmission through the
inelastic sidebands is due to destructive quantum interference
between these two amplitudes. At normal incidence
$k_{n}-k_{n-1}=q_{n-1}-q_{n}=\omega/v_{f}$ and
 $k_{-n}-k_{-(n-1)}=q_{-(n-1)}-q_{-n}=-\omega/v_{f}$ with $\left(n=1,2,...\right)$, result from the gapless
and linear electron dispersion law near the Fermi energy. Therefore, we can also write the
expression (\ref{condinterferencia}) in general form as:
\begin{equation}
\delta_{c}+L_{2}+L_{1}=\pi,\label{condgral}
\end{equation}
where $L_{1}$ and $L_{2}$ are expressed in units of $v_{f}/\omega$
and the critical phase difference depends only on barrier and well
width. Thus, Eq.(\ref{condgral}) indicates a simultaneous
destructive interference of contiguos channels at $\phi_{0}=0$. Fig. 3a) shows the transmission probability through the elastic
channel as a function of $L_{1}$ and $\delta$ at $\alpha_{1}=7$ calculated after Eq.(\ref{channeltra}). The straight line in this figure is Eq.(\ref{condgral}) at
$L_{2}=0.25$. In panel b) only the region where $1-T_{0}<0.005$
for different well widths is shown. The analytical result presented
as condition (\ref{condgral}) remains valid even in the regime
$\alpha_{1}\gg1$, when a large number of channels coexist, and show
excellent agreement with numerical solutions in Fig. 3. Thus, quantum
interference between first order process is the physical basis
behind Eq.(\ref{condgral}).
\begin{figure}[bp]
\begin{centering}
\hspace*{-.5cm}\includegraphics[scale=0.5]{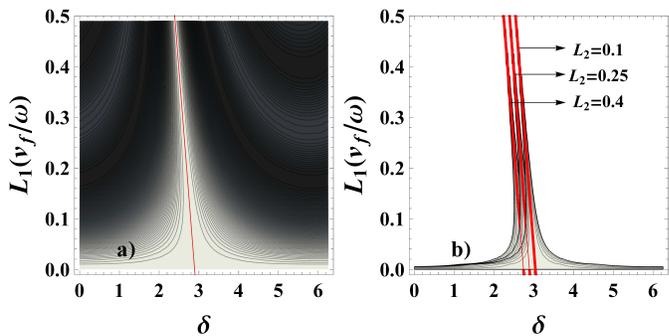}
\par\end{centering}
\caption{(Color online) $T_{0}$ as a function of $L_{1}$ and $\delta$ at $\phi_{0}=0$,
$\alpha_{1}=7$ and a) $L_{2}=0.25$ b) well widths; $0.1$,
$0.25$ and $0.4$. The red line is (\ref{condgral})
at $L_{2}$ fixed.}
\end{figure} Now, it is clearly seen why
$\left|\alpha_{1}-\alpha_{3}\right|<1$ is necessary for the
suppression of the transmission through inelastic sidebands. The
same amplitude of probability for the process of absorption
(emission) on both barrier region is required to achieve destructive
interference. It should be pointed out that, at normal incidence,
perfect total transmission, reported for static~\cite{13} and
oscillating~\cite{15} single barriers
persists for harmonically driven DBS regardless of interference effects.\\
\textbf{Oblique incidence.}\\
Now, resonant tunneling through DBS at oblique incidence under time
periodic field is investigated. Due to the existence of evanescent
modes inside the static barrier~\cite{14}, the transmission, as a function of
the incident energy, has a gap which can be tuned changing the
height of the barrier and/or the angle of incidence~\cite{18}.
Because of coupling of evanescent states in the barrier with
resonant states in the well, several resonance peaks with a unity
value appear in the transmission gap~\cite{19}. In the following,
the influence of oscillation amplitudes and interference effects on
resonant tunneling is studied. Fig. 4 shows total transmission
probability as a function of the energy of incident electrons at
$\phi_{0}=\pi/18$ for different amplitudes of the oscillating
field and $\delta=0$. In Fig. 4a) the transmission probability for
the static double barrier is shown as it corresponds to
$\alpha_{1}=\alpha_{3}=0$, which was previously obtained. Resonant
peaks are narrow and could have very important applications in
high-speed devices based on graphene as has been suggested in
Ref.~\cite{19}. Considering the effect of PAT, some satellite peaks
appear on both sides of the two main resonant peaks, and as the
amplitude of the oscillating field increases, satellite peaks move
away from the two main resonant peaks due to emission or absorption
of a greater number of photons. Coupling between evanescent wave
inside the barriers and propagating wave inside the well occur for
several energy values.
\begin{figure}[tbp]
\hspace*{-.75cm}\includegraphics[scale=0.55]{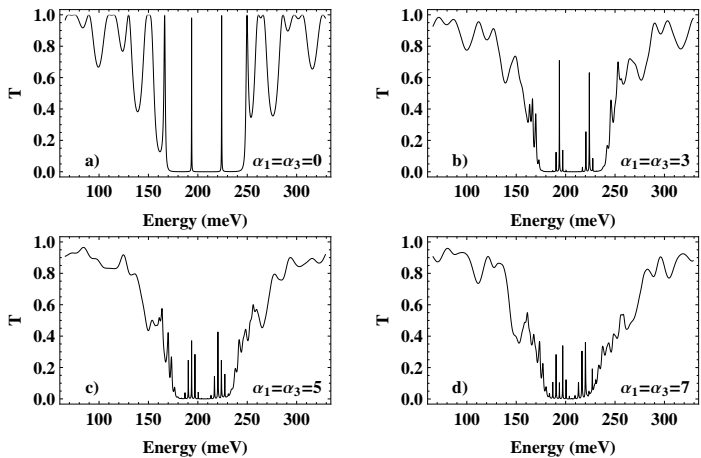}
\caption{Total transmission probability on incident energy at $\phi_{0}=\pi/18$ and $\delta=0$ for several value of the modulation amplitude.}
\label{fig4}
\end{figure}
The positions of the satellite peaks are $E_{1,2}\pm n\hbar\omega$
where $E_{1,2}$ are the positions of the main resonant peaks. At
$\alpha_{1}=\alpha_{3}=3$ only two main peaks with few small
satellites are obtained. However, at high amplitude modulation
($\alpha_{1}=\alpha_{3}=7$), the two main peaks lose its dominance and the intensities of main peaks and satellite peaks are
reversed. Therefore, strong quenching of resonant transmission with
increasing amplitude of oscillating field is found as shown in panel
d) of Fig. 4.\\
For non-normal incidence $k_{\pm n}-k_{\pm(n-1)}$ and
$q_{\pm(n-1)}-q_{\pm n}$ depend on $E\pm n\hbar\omega$, $V_{0}$ and
$\phi_{0}.$ Then, it is not possible to determine analytically a
critical phase difference which cancels simultaneously all contiguos
channels. However, suppression of inelastic sidebands at
$\delta=2.64$ is obtained numerically not only for normal incidence
but $\left(T_{0}>0.95\, T\right)$ for different values of the
incident energy and incident angle at $\alpha_{1}=\alpha_{3}=7$ as
shown in Fig. 5.
\begin{figure}[tbp]
\hspace*{-.5cm}\includegraphics[scale=0.45]{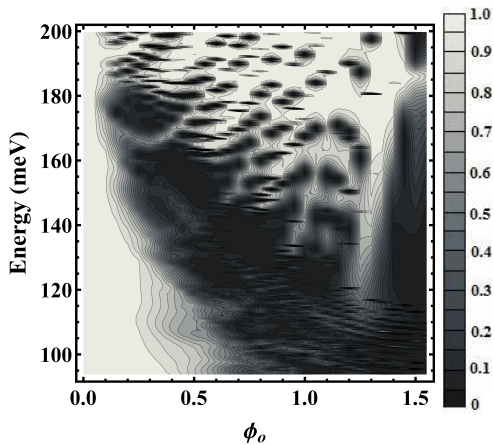}
\includegraphics[scale=0.76]{legendaver2.eps}
\caption{$T_{0}/T$ as a function of incident energy and angle for $\alpha_{1}=\alpha_{3}=7$ and $\delta=2.64$.}
\label{fig5}
\end{figure}
Thus, destructive interference effects are robust, since the phase
diagram ($E$ vs $\phi_{0}$) in Fig. 5 exhibits many white zones at
oblique incidence, corresponding to strong quenching of the
transmission through inelastic sidebands. These zones change with phase difference and become highly anisotropic due to the
chiral nature of quasiparticles in graphene. Fig. 6 shows $T_{0}/T$ as a function of
$\delta$ for incident energy corresponding to first main resonant
peaks ($E_{1}=193.695$ $meV$) and $\phi_{0}=\pi/18$. Note that transmission is practically only through central band for
$\delta=2.76$ and for other phase differences inelastic sidebands
will become more important. Thus, changing $\delta$ we shall observe
radical influence in the resonant features.
\begin{figure}[tbp]
\includegraphics[scale=0.47]{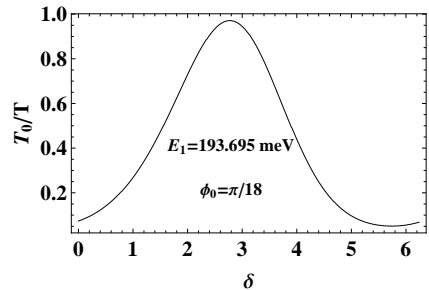}
\caption{$T_{0}/T$ as a function of phase difference at $\alpha_{1}=\alpha_{3}=7$.}
\label{fig6}
\end{figure}
Fig. 7 shows total transmission probability as a function of energy
at $\phi_{0}=\pi/18$ and $\alpha_{1}=\alpha_{3}=7$ for two values of
the phase difference. Adjusting the value of $\delta$ resonant
transmission can be controlled. For $\delta=2.76$ satellite peaks
are strongly suppressed and almost perfect transmission is obtained
for $E_{1}=193.695$ $meV$ whereas at $\delta=0$ a large number of
sidebands coexist and resonant tunneling practically disappears.
\begin{figure}[tbp]
\includegraphics[scale=0.44]{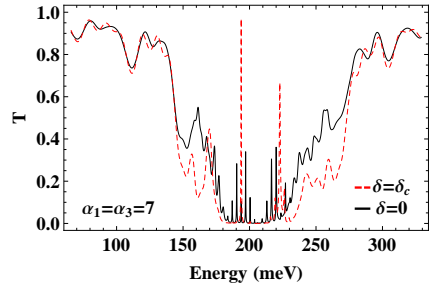}
\caption{(Color online) $T$ as a function of incident energy at $\phi_{0}=\pi/18$ for $\delta=0$ (black line) and $\delta=2.76$ (dashed red line).}
\label{fig7}
\end{figure}
Thus, quantum interference makes a considerable contribution
to total transmission at non-normal incidence.\\
\textbf{Oscillating well.}\\
In addition, total transmission probability has been also calculated
when the harmonic potential is applied at the well instead of the
barrier. In Fig. 8 PAT in a driven DBS with oscillating quantum well
(panel b) and oscillating quantum barriers (panel a), are compared.
\begin{figure}[tbp]
\hspace*{-.75cm}\includegraphics[scale=0.55]{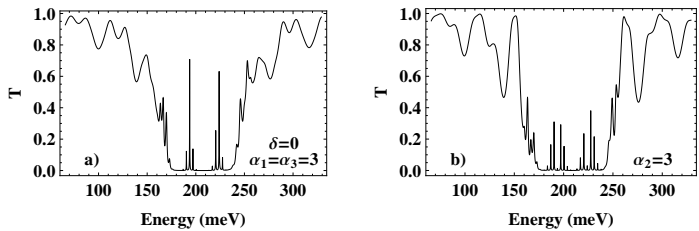}
\caption{$T$ as a function of incident energy at $\phi_{0}=\pi/18$
for a) oscillating quantum barriers and b) oscillating quantum well.} \label{fig8}
\end{figure}
We observe that the driving field with amplitude $3\hbar\omega$
is more effective in quenching of resonant tunneling when applied to the well. It occurs because, for the incident energy
in the gap region $(V_{0}-\hbar k_{y} v_{f}<E<V_{0}+\hbar k_{y} v_{f})$, the evanescent modes appear inside the barriers and the electrons
reflect back and forth several rounds in the well. Thus, for oscillating quantum well the electrons have enough time
to interact with the driving field contrary to oscillating quantum barriers. Consequently, resonant peaks
are more strongly quenched in driven well.\\
In summary, we have carried out a study of electron PAT through
graphene-based symmetric double barriers driven by a periodic
potential. Barriers oscillate with the same frequency $\omega$,
different amplitudes and phase difference $\delta$. The
time-periodic electromagnetic field generates additional sidebands
at energies $E\pm n\hbar\omega$ in the transmission probability due
to photon absorption or emission. At normal incidence, perfect total transmission probability (Klein tunneling) persist for harmonically driven DBS.
A critical phase difference is found such that, total
suppression of inelastic sidebands due to destructive interference
between waves is obtained. A condition for the simultaneous cancellation
of contiguos channels as a result of linear dispersion law, is derived for small amplitude modulation
and is valid even in the regime ($\alpha_{1}=\alpha_{3}\gg1$). Thus, energy of the
transmitted electrons is sensitive to phase difference and inelastic sidebands can be removed by tuning structural parameters of the DBS
according to Eq.(\ref{condgral}). For oblique incidence, suppression of inelastic sidebands
occurs and destructive interference plays a fundamental role in the total transmission probability. Resonant
transmission may be regulated, as resonant
tunneling is quenched at $\delta=0$ and recovered for certain phase
difference. Moreover, quenching of resonant transmission in driven well is more drastic than driven barriers.
In conclusion, we have shown that quantum interference
has an important effect on quasiparticles tunneling through a
time-dependent graphene-based double barrier. This phenomenon
has potential applications in graphene-based electronic
devices such as energy filters and high-frequency radiation
detectors.




\end{document}